\documentclass[12pt,a4paper]{article}
\usepackage{epsfig}
\begin{document}
\title{Detecting the neutral top-pion at future muon colliders}
\author{Chongxing Yue$^{a}$, Lanjun Liu$^{b}$, Dongqi Yu$^{a}$\\
{\small a: Department of Physics, Liaoning  Normal University,
Dalian 116029. P.R.China}
\thanks{E-mail:cxyue@lnnu.edu.cn}
\\ {\small b:College of Physics and Information Engineering,}\\
\small{Henan Normal University, Xinxiang  453002. P.R.China}
 }
\date{\today}
\maketitle

\begin{abstract}
\hspace{5mm}The signals of the neutral top-pion $\pi^{0}_{t}$ at
the first muon collider (FMC) are discussed by calculating its
contributions to the processes $\mu^{+}\mu^{-}\longrightarrow
b\bar{b}$ and $\mu^{+}\mu^{-}\longrightarrow \bar{t}c$. We find
that the contributions to  the process
$\mu^{+}\mu^{-}\longrightarrow b\bar{b}$ are very small and the
ratio of signal over square root of the background ($S/\sqrt{B})$
is smaller than $0.2$. However, in most of the parameter space,
the $\pi^{0}_{t}$ can generate $36$ and up to $649$ observable
$\bar{t}c$ events at the FMC. The signals of the neutral top-pion
$\pi^{0}_{t}$ may be detected via the process
$\mu^{+}\mu^{-}\longrightarrow \bar{t}c$ at the FMC.
\end {abstract}

\vspace{2.0cm} \noindent
 {\bf PACS number(s)}:12.60Nz,14.80.Mz,12.15.Lk,14.65.Ha

\newpage

\vspace{.5cm} \noindent{\bf I. Introduction}

A muon collider is an excellent tool to study the properties of a
heavy scalar or pseudoscalar [1,2]. It is thought that the first
muon collider (FMC) can achieve the same integrated luminosity and
energy as a high energy $e^{+}e^{-}$ collider. The FMC can
 explore all the same physics that is accessible at an
$e^{+}e^{-}$ collider of the same energy. Furthermore, it has been
shown that a large number of Higgs bosons [1] or new particles,
such as technihadrons and technipions [3,4], can be produced
through an s-channel resonance process and the flavor changing
scalar (FCS) couplings [5] can be tested at the FMC. Thus, the FMC
makes possible precise measurements of the total widths and masses
of the various neutral particles, which will open a window towards
the new physics.

Until a Higgs boson with large coupling to a gauge boson pair is
discovered, the possibility of the electroweak symmetry broken by
new strong interactions still exists [6]. The most commonly
studied class of theories is technicolor (TC) [7], which
dynamically breakes the electroweak symmetry. Although TC models
have many theoretical problems as well as conflict with data and
broad classes of these models have been ruled out, there are still
viable models worthy of investigation in light of the capabilities
of the current collider experiments. Topcolor-assisted technicolor
(TC2) models [8] are such type of examples.

The common feature of all TC2 models is that topcolor interactions
generate the main part of large top quark mass and only make small
contributions to electroweak symmetry breaking (EWSB). In order
that topcolor interactions are natural, i.e. without introducing
large isospin violation, it is necessary that EWSB is still mainly
generated by TC interactions. For TC2 models, ETC interactions are
still needed to generate the masses of light quarks and contribute
a few $GeV$ to top quark mass, i.e. $\varepsilon m_{t}$ with
$\varepsilon\ll1$ [8]. Thus, the presence of a number of pseudo
Goldstone bosons (PGB's), including the technipions in TC sector
and three top-pions ($\pi^{\pm,0}_{t}$) in topcolor sector, in the
low-energy spectrum is an inevitable feature of these models.
These new particles are most directly related to EWSB. Thus,
studying the possible signatures of these new particles at present
and future high energy experiments would provide crucial
information for EWSB and fermion flavor physics as well.

The production and decay of the technipions and top-pions have
been extensively studied in several instances [9,10]. Combining
resonance and non-resonance contributions, the signals of
technipions are recently studied at lepton colliders and  hadron
colliders [11]. Ref.[12] has discussed the single production of
the neutral top-pion $\pi^{0}_{t}$ at hadron colliders and high
energy $e^{+}e^{-}$ colliders (LCs). The observability of the
neutral top-pion $\pi_{t}^{0}$ and the charged top-pions
$\pi_{t}^{\pm}$ has been studied via considering the single top
production at hadron colliders  and linear colliders [13,14].

The effects of the top-pion on physical observables are governed
by its mass $m_{\pi_{t}}$, while the large couplings of top-pions
to quarks and to gauge bosons are to large degree
model-independent [15]. Thus, if the neutral top-pion
$\pi_{t}^{0}$ is discovered at the LHC or other future collider
experiments, it is need to consider its resonance production at
the FMC, which can make possible precise measurements of the
couplings and mass of the neutral top-pion $\pi_{t}^{0}$. In this
paper, we consider the s-channel resonance production of the
neutral top-pion $\pi^{0}_{t}$ at the FMC with the centre-of-mass
energy $\sqrt{s}=200GeV-500GeV$ and explore the potential of the
FMC for detecting this new particle.

This paper is organized as follows: in section 2, we give the
possible couplings of the neutral top-pion $\pi^{0}_{t}$ to the
ordinary particles. The resonance production cross section
$\sigma(b\bar{b})$ of the process $\mu^{+}\mu^{-}\longrightarrow
\pi^{0}_{t}\longrightarrow b\bar{b}$ and the ratio of signal over
square root of the background $(S/\sqrt{B})$ are calculated in
section 3. For TC2 models, the topcolor interactions are
non-universal, the neutral top-pion $\pi^{0}_{t}$ has large FCS
coupling $\pi^{0}_{t}\bar{t}c$, the possibility of detecting
$\pi^{0}_{t}$ via the process $\mu^{+}\mu^{-}\longrightarrow
\pi^{0}_{t}\longrightarrow \bar{t}c$ is studied in section 3. Our
conclusions are given in section 4.

\vspace{0.5cm} \noindent{\bf II The couplings of the neutral
top-pion $\pi^{0}_{t}$ to ordinary particles}

For TC2 models [8], TC interactions play a main role in breaking
the electroweak symmetry. Topcolor interactions make small
contributions to EWSB, and give rise to the main part of the top
quark mass, $(1-\varepsilon)m_{t}$, with the parameter
$\varepsilon\ll1$. Thus, there is the following relation:
\begin{equation}
\nu^{2}_{\pi}+F^{2}_{t}=\nu^{2}_{W},
\end{equation}
where $\nu_{\pi}$ represents the contributions of TC interactions
to EWSB, $\nu_{W}=\nu/\sqrt{2}=174GeV$. $F_{t}=50GeV$ is the
physical top-pion decay constant, which can be estimated from the
Pagels-Stokar formula. This means that the masses of gauge bosons
W and Z are given by absorbing the linear combination of the
top-pions and technipions. The orthogonal combination of the
top-pions and technipions remains unabsorbed and physical.
However, the absorbed Goldstone linear combination is mostly the
technipions while the physical linear combination is mostly the
top-pions, which are usually called physical top-pions
($\pi^{\pm}_{t},\pi^{0}_{t}$). The existence of the physical
top-pions in the low-energy spectrum can be seen as
characteristics of topcolor scenario, regardless of the dynamics
responsible for EWSB and other quark masses [15].

For TC2 models, the underlying interactions, topcolor
interactions, are non-universal and therefore do not posses a GIM
mechanism. The non-universal gauge interactions result in the new
FC coupling vertices when one writes the interactions in the quark
mass eigen-basis. Thus, the top-pions have large Yukawa couplings
to the third generation quarks and can induce the new FCS
couplings. The couplings of the neutral top-pion $\pi^{0}_{t}$ to
the third generation quarks including the $t-c$ transition can be
written as [8,14]:
\begin{equation}
\frac{m_{t}}{\sqrt{2}F_{t}}\frac{\sqrt{\nu^{2}_{W}-F^{2}_{t}}}
{\nu_{W}}[K^{tt}_{UR}K^{tt}_{UL}\bar{t}_{L}t_{R}\pi^{0}_{t}
+\frac{m_{b}-m'_{b}}{m_{t}}\bar{b_{L}}b_{R}\pi^{0}_{t}+K^{tc}_{UR}
K^{tt}_{UL}\bar{t_{L}}c_{R}\pi^{0}_{t}+h.c.],
\end{equation}
where $m'_{b}$ is the ETC generated part of the bottom-quark mass.
Similar to Ref.[10], we take $m'_{b}=0.1\varepsilon m_{t}$.
$K^{tt}_{UL}$ is the matrix element of the unitary matrix $K_{UL}$
which the CKM matrix can be derived as $V=K^{-1}_{UL}K_{DL}$ and
$K^{ij}_{UR}$ are the matrix elements of the right-handed rotation
matrix $K_{UR}$. Their values can be written as [14] :
\begin{equation}
K^{tt}_{UL}=1,\hspace{5mm}K^{tt}_{UR}=1-\varepsilon,\hspace{5mm}
K^{tc}_{UR}\leq\sqrt{2\varepsilon-\varepsilon^{2}}.
\end{equation}
In the following calculation, we will take
$K^{tc}_{UR}=\sqrt{2\varepsilon-\varepsilon^{2}}$ and take the
parameter $\varepsilon$ as a free parameter.

The couplings of the neutral top-pion $\pi^{0}_{t}$ to the first,
second generation fermions and the third generation leptons can be
written as :
\begin{equation}
\frac{m_{f}}{\sqrt{2}F_{t}}\frac{F_{t}}{\nu_{W}}\bar{f}
\gamma^{5}f\pi^{0}_{t}=\frac{m_{f}}{\nu}\bar{f}\gamma^{5}f\pi^{0}_{t}.
\end{equation}

The neutral top-pion $\pi^{0}_{t}$, as an isospin-triplet, can
couple to a pair of gauge bosons through the top quark trangle
loop in an isospin violating way similar to the couplings of QCD
pion $\pi^{0}$ to a pair of gauge bosons. The relevant formula of
these couplings has been given in Ref.[10].

Ref.[8] has estimated the mass of the top-pion in the fermion loop
approximation and given $180GeV\leq m_{\pi_{t}}\leq240GeV$ for
$m_{t}=175GeV$ and $0.03\leq \varepsilon \leq 0.1$. The limits on
the mass of the top-pion may be obtained via studying its effects
on various experimental observables. For example, Ref.[16] has
shown that the process $b\longrightarrow s\gamma$, $B-\bar{B}$
mixing and $D-\bar{D}$ mixing demand that the top-pions are likely
to be light, with masses of the order of a few hundred $GeV$.
Since the negative top-pion corrections to the $Z\longrightarrow
b\bar{b}$ branching ratio $R_{b}$ become smaller when the top-pion
is heavier, the LEP-SLD data of $R_{b}$ give rise to a certain
lower bound on the top-pion mass [15]. It was shown that the
top-pion mass should not be lighter than the order of $1TeV$ to
make TC2 models consistent with the LEP-SLD data [17]. We
restudied the problem in Ref.[18] and find that the top-pion mass
$m_{\pi_{t}}$ is allowed to be in the range of a few hundred $GeV$
depending on the models. Thus, the value of the top-pion mass
$m_{\pi_{t}}$ remains subject to large uncertainty [7].
Furthermore, Refs.[13,14] have shown that the top-pion mass
$m_{\pi_{t}}$ can be explored up to $300-350GeV$ via the processes
$p\bar{p}\longrightarrow\pi_{t}^{0}\longrightarrow\bar{t}c$ and
$p\bar{p}\longrightarrow\pi_{t}^{\pm}x$ at the Tevatron and LHC.
Thus, we will take $m_{\pi_{t}}$ as a free parameter and assume it
to vary in the range of $200GeV-400GeV$ in this paper. In this
case, the possible decay modes of $\pi^{0}_{t}$ are
$b\bar{b},\bar{t}c,\bar{f}f$($f$ is the first or second generation
fermions or the third generation leptons), $t\bar{t}$(if
kinematically allowed), $gg$, $\gamma\gamma$, and $Z\gamma$. The
total decay width of $\pi^{0}_{t}$ can be written as:
\begin{eqnarray}
\Gamma_{\pi_{t}}&=&\Gamma(\pi^{0}_{t}\longrightarrow
b\bar{b})+\Gamma(\pi^{0}_{t}\longrightarrow
\bar{t}c)+\Gamma(\pi^{0}_{t}\longrightarrow
f\bar{f})+\Gamma(\pi^{0}_{t}\longrightarrow gg) \nonumber\\
&+&\Gamma(\pi^{0}_{t}\longrightarrow
\gamma\gamma)+\Gamma(\pi^{0}_{t}\longrightarrow
Z\gamma)+\Gamma(\pi^{0}_{t}\longrightarrow t\bar{t}) (
m_{\pi_{t}}\geq2m_{t}).
\end{eqnarray}

\vspace{0.5cm}
\noindent{\bf III The signals of the neutral
top-pion $\pi^{0}_{t}$ at the FMC}

From above discussions, we can see that the neutral top-pion
$\pi^{0}_{t}$ can be produced at the FMC, operating at a
centre-of-mass energy $\sqrt{s}$ of up to $500GeV$. In spite of
the fact that the  coupling $\pi^{0}_{t}\mu^{+}\mu^{-}$, being
proportional to $m_{\mu}/\nu$, is very small, if the FMC
 run on the $\pi^{0}_{t}$ resonance
$(\sqrt{s}=m_{\pi_{t}})$, the $\pi^{0}_{t}$ may be produced at an
appreciable rate. For $m_{\pi_{t}}\leq2m_{t}$, the main decay
modes of $\pi^{0}_{t}$ are $\bar{t}c,b\bar{b}$ and $gg$. So, in
this section, we will consider the possibility of detecting
$\pi^{0}_{t}$ via the processes
$\mu^{+}\mu^{-}\longrightarrow\pi^{0}_{t}\longrightarrow b\bar{b}$
and $\mu^{+}\mu^{-}\longrightarrow \pi^{0}_{t}\longrightarrow
\bar{t}c$ at the FMC.

\vspace{0.5cm}
\noindent{\bf 1.The neutral top-pion $\pi^{0}_{t}$
and the process $\mu^{+}\mu^{-}\longrightarrow b\bar{b}$}

Convoluted with the collider energy distribution, the s-channel
resonance cross section for production of a final state $x$
generated by the neutral top-pion $\pi^{0}_{t}$ at the FMC is
given by [1]:
\begin{equation}
\sigma(x)\approx\frac{4\pi}{m^{2}_{\pi_{t}}}\frac{B_{r}(\pi^{0}_{t}
\longrightarrow\mu^{+}\mu^{-})B_{r}(\pi^{0}_{t}\longrightarrow x)}
{[1+\frac{8}{\pi}(\frac{\sigma_{\sqrt{s}}}{\Gamma_{\pi_{t}}})^{2}]^{1/2}}.
\end{equation}
The Gaussian spread in the  beam energy $\sqrt{s}$ is given by
$\sigma_{\sqrt{s}}=\frac{R}{\sqrt{2}}\sqrt{s}$. The energy
resolution $R$  of each beam is expected to be in the range of
$0.003\%-0.05\%$ and we will take $R=0.03\%$.

In Fig.1, we plot the resonance production section
$\sigma(b\bar{b})$ versus the top-pion mass $m_{\pi_{t}}$ for
$\sqrt{s}=m_{\pi_{t}}$ and three values of the parameter
$\varepsilon$. We can see from Fig.1 that $\sigma(b\bar{b})$
decreases with the top-pion mass $m_{\pi_{t}}$ and the parameter
$\varepsilon$ increasing. The value of $\sigma(b\bar{b})$ is in
the range of 2.1fb---0.01fb for $0.02\leq\varepsilon\leq0.08$ and
$200GeV\leq m_{\pi_{t}}\leq350GeV$. Thus, there may be 1---42
$b\bar{b}$ events to be produced at the FMC with
$\sqrt{s}=200GeV-500GeV$ and a yearly integrated luminosity of
$L=20fb^{-1}$. To see whether the neutral top-pion $\pi^{0}_{t}$
can be detected via the process $\mu^{+}\mu^{-}\longrightarrow
b\bar{b}$ at the FMC, it is need to further calculate the ratio of
signal over square root of the background $\frac{S}{\sqrt{B}}$.

To solve the phenomenological difficulties of the traditional TC
models [7], TC2 models [8] were proposed by combining technicolor
interactions with  topcolor interactions for the third generation
quarks. TC2 models predict a number of technipions in the TC
sector. Refs.[3,4] have pointed out that the color-singlet
technipion $P^{0}$ can be significantly produced in the s-channel.
Thus, the backgrounds of the process
$\mu^{+}\mu^{-}\longrightarrow \pi^{0}_{t}\longrightarrow
b\bar{b}$ mainly come from the exchange of the gauge bosons
$\gamma, Z$ and of the color-singlet technipion $P^{0}$ in the
s-channel. We calculate the value of $\frac{S}{\sqrt{B}}$ and find
that it is smaller than 0.2 in all of the parameter space. So the
neutral top-pion $\pi^{0}_{t}$ can not be detected via the process
$\mu^{+}\mu^{-}\longrightarrow b\bar{b}$ at the FMC.

\begin{figure}[h]
\begin{center}
\begin{picture}(200,100)(0,0)
\put(-100,-180){\epsfxsize130mm\epsfbox{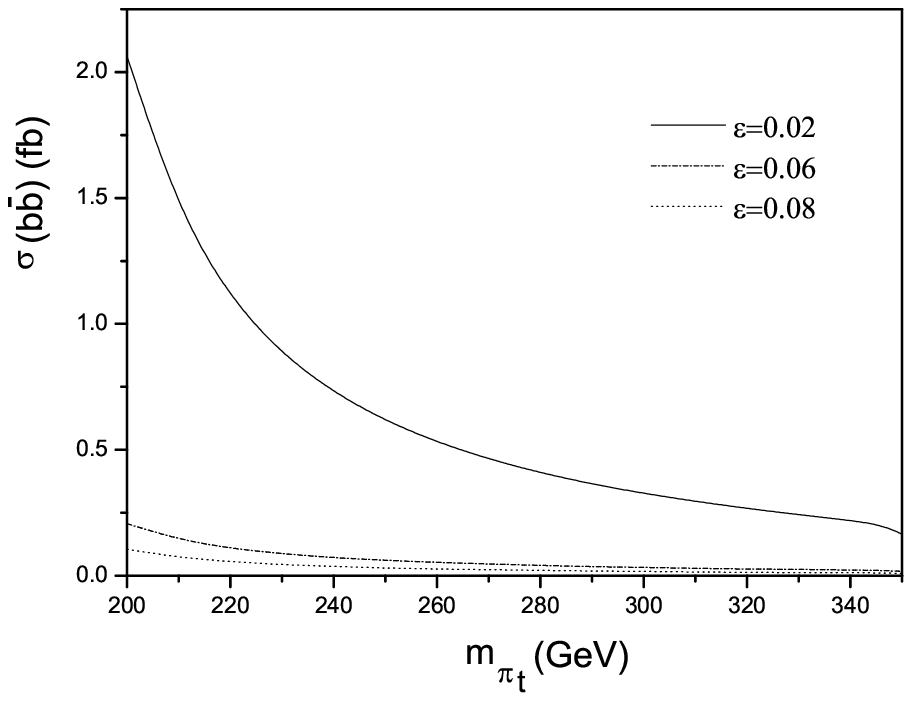}}
\put(-120,-180){Fig.1\hspace{5mm}The resonance production cross
section $\sigma(b\bar{b})$ versus the top-pion mass $m_{\pi_{t}}$}
\put(-80,-200){for three values of the parameter $\varepsilon$.}
\end{picture}
\end{center}
\end{figure}

\vspace{6.cm}

\vspace{0.5cm} \noindent{\bf 2. The neutral top-pion $\pi^{0}_{t}$
and the process $\mu^{+}\mu^{-}\longrightarrow \bar{t}c$}

From section 2 we can see that the most dominant decay mode of the
neutral top-pion $\pi^{0}_{t}$ is $\bar{t}c$ for
$m_{\pi_{t}}\leq350GeV$. Thus, compared to the process
$\mu^{+}\mu^{-}\longrightarrow b\bar{b}$, $\pi^{0}_{t}$ can give
significant contributions to the process
$\mu^{+}\mu^{-}\longrightarrow \bar{t}c$ via the $\pi^{0}_{t}$
exchange in the s-channel. Furthermore, it is well known that
there is no flavor changing neutral current (FCNC) at tree-level
in the SM. The production cross sections of the FCNC process are
very small at one-loop level due to the unitary of CKM matrix. The
FCNC processes can be used to search for new physics. Any
observation of the FC coupling deviated from that in the SM would
unambiguously signal the presence of new physics. Thus, the
process $\mu^{+}\mu^{-}\longrightarrow \pi^{0}_{t}\longrightarrow
\bar{t}c$ will give a signal which should be easy to identify. The
neutral top-pion $\pi^{0}_{t}$ may be easy detected via this
process at the FMC.

In Fig.2 we show the resonance production cross section
$\sigma(\bar{t}c)$ as a function of the top-pion mass
$m_{\pi_{t}}$ for $\sqrt{s}=m_{\pi_{t}}$ and three values of the
parameter $\varepsilon$. We can see from Fig.2 that the
$\sigma(\bar{t}c)$ decreases with  $m_{\pi_{t}}$ and the parameter
$\varepsilon$ increasing. For $m_{\pi_{t}}\geq2m_{t}$, the cross
section $\sigma$ drops considerably since the $t\bar{t}$ channel
opens up and the branching ratio $B_{r}
(\pi^{0}_{t}\longrightarrow \bar{t}c)$ drops substantially. The
value of the $\sigma(\bar{t}c)$ is larger than 3.7fb for
$\varepsilon\leq0.08$ and $200GeV\leq m_{\pi_{t}}\leq350GeV$.
Thus, the FMC with $\sqrt{s}=200GeV-500GeV$ and a yearly
integrated luminosity of $L=20fb^{-1}$ will generate tens and up
to thousand $\bar{t}c$ events for $200GeV\leq
m_{\pi_{t}}\leq350GeV$ and $0.02\leq\varepsilon\leq0.08$. For
example, it will generate 280 $\bar{t}c$ events per year for
$m_{\pi_{t}}=300GeV$ and $\varepsilon=0.06$. Thus, we can study
the signals and observability of the neutral top-pion
$\pi^{0}_{t}$ via the process $\mu^{+}\mu^{-}\longrightarrow
\bar{t}c$ at the FMC.
\begin{figure}[h]
\begin{center}
\begin{picture}(200,100)(0,0)
\put(-100,-180){\epsfxsize130mm\epsfbox{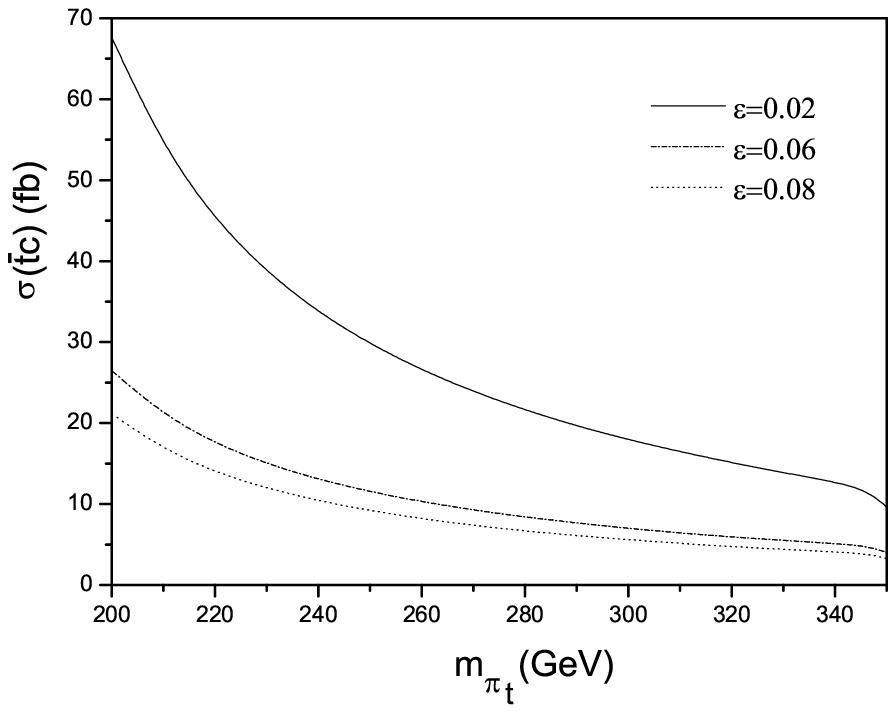}}
\put(-120,-180){Fig.2\hspace{5mm}The resonance production cross
section $\sigma(\bar{t}c)$ as a function of $m_{\pi_{t}}$ }
\put(-80,-200){for three values of the parameter $\varepsilon$.}
\end{picture}
\end{center}
\end{figure}
\vspace{6.5cm}

The parameter $\varepsilon$ of TC2 models represents the
contributions of ETC interactions or other interactions to the
mass of the top quark and is in the range of 0.02---0.1 [8]. The
current constraints on the free parameter $\varepsilon$ from low
energy data (such as the $D-\bar{D}$ and $B-\bar{B}$ mixing and
the $b\rightarrow s\gamma$ rate) are rather weak. To see the
effects of the parameter $\varepsilon$ on the resonance production
cross section $\sigma(\bar{t}c)$, we plot $\sigma(\bar{t}c)$ as a
function of the parameter $\varepsilon$ for
$\sqrt{s}=m_{\pi_{t}}=300GeV$ in Fig.3. We can see from Fig.3 that
the cross section $\sigma(\bar{t}c)$ is larger than 5.6fb for
$\varepsilon\leq0.08$ and the number of yearly generated
$\bar{t}c$ events is larger than 110.

\begin{figure}[h]
\begin{center}
\begin{picture}(200,100)(0,0)
\put(-80,-180){\epsfxsize130mm\epsfbox{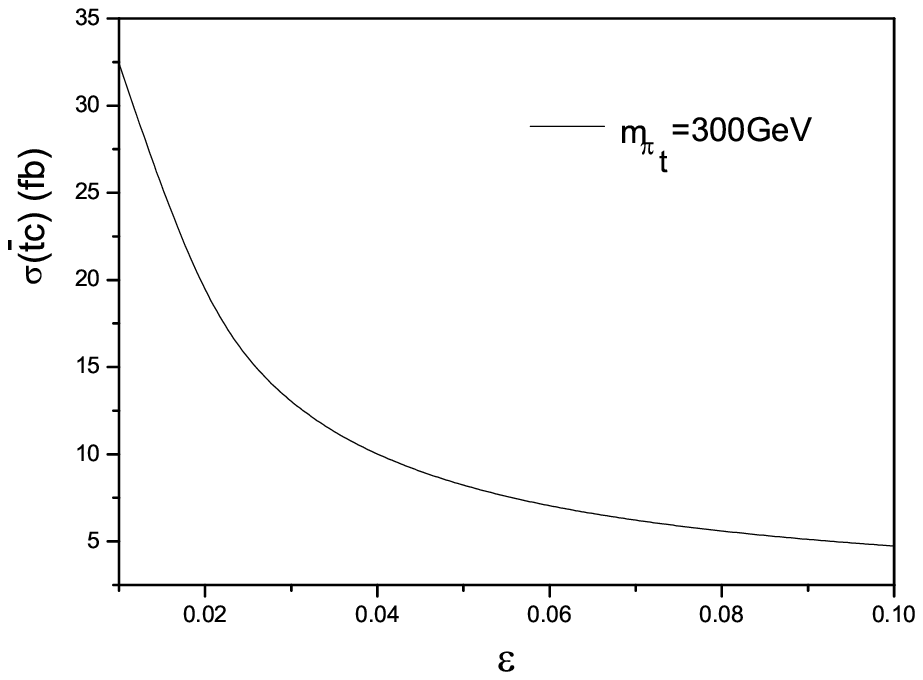}}
\put(-120,-180){Fig.3\hspace{5mm}The resonance production cross
section $\sigma(\bar{t}c)$ as a function of the}
\put(-80,-200){parameter $\varepsilon$ for
$\sqrt{s}=m_{\pi_{t}}=300$.}
\end{picture}
\end{center}
\end{figure}
\vspace{6.5cm}

The possible observable final states of the FCNC process
$\mu^{+}\mu^{-}\longrightarrow\bar{t}c$ are $\bar{b}cj_{1}j_{2}$,
where $j_{1}$ and $j_{2}$ are light jets coming from
$t\longrightarrow bW^{+}$ followed by $W^{+}\longrightarrow
u\bar{d}$ or $c\bar{s}$, and $\bar{b}cl^{+}\nu_{l}$, which come
from $t\longrightarrow bW^{+}\longrightarrow bl^{+}\nu_{l}$ with
$l=e, \mu$ or $\tau$. These two final states occur with branching
of $2/3$ and $1/3$, respectively. The leading SM backgrounds of
the FCNC process $\mu^{+}\mu^{-}\longrightarrow\bar{t}c$ mainly
come from $W$ pair production and from $W$ bremsstrahlung in
$\mu^{+}\mu^{-}\longrightarrow W+2-$jets. The techniques for
suppressing this kind of backgrounds were discussed in Ref.[19].

We define the background-free observable cross section
$\overline{\sigma(\bar{t}c)}$ as the effective cross section
including b-tagging efficiency ($\varepsilon_{b}$) and top quark
reconstruction efficiency ($\varepsilon_{t}$):
$\overline{\sigma(\bar{t}c)}=\varepsilon_{b}\varepsilon_{t}\sigma(\bar{t}c)$.
In order to estimate the number of the observable events, we
assume $\varepsilon_{b}=60\%$ and $\varepsilon_{t}=80\%$ as done
in Ref.[19]. Then the yearly production observable $\bar{t}c$
events at the FMC can be easily estimated. The FMC with
$\sqrt{s}=200GeV-500GeV$ and yearly integrated luminosity of
$L=20fb^{-1}$ will generate 36 and up to 649 observable $\bar{t}c$
events for $200GeV\leq m_{\pi_{t}}\leq350GeV$ and
$0.02\leq\varepsilon\leq0.08$. Thus, it is easy to detect the
neutral top-pion $\pi_{t}^{0}$ via the process
$\mu^{+}\mu^{-}\longrightarrow\pi_{t}^{0}\longrightarrow\bar{t}c$.
We can study the properties of the neutral top-pion $\pi_{t}^{0}$
via the process $\mu^{+}\mu^{-}\longrightarrow\bar{t}c$ at the
FMC.

\vspace{0.5cm} \noindent{\bf IV Conclusions}

The top quark, with a mass of the order of the electroweak scale,
is singled out to play a key role in the dynamics of EWSB and
flavor symmetry breaking. There may be a common origin for EWSB
and top quark mass generation. Much theoretical work has been
carried out in connection to the top quark and EWSB. Various kinds
of strong top dynamical models have been proposed, including TC2
models [8], flavor-universal TC2 models [20], top see-saw models
[21], and top flavor see-saw models [22]. The common feature of
such type of models is that topcolor interactions generate the
main part of the top quark mass and also make small contributions
to EWSB. EWSB is mainly generated by TC interactions or other
interactions. Then, the presence of physical top-pions in the low
energy spectrum is an inevitable feature of these models. Thus,
studying the production and decay of these new particles at
present or future high energy colliders is of special interest. It
will be helpful to test the topcolor scenario and understand EWSB
mechanism.

In this paper, we have studied the possible for detecting the
neutral top-pion $\pi^{0}_{t}$ at the FMC via the processes
$\mu^{+}\mu^{-}\longrightarrow \pi^{0}_{t}\longrightarrow
b\bar{b}$ and $\mu^{+}\mu^{-}\longrightarrow
\pi^{0}_{t}\longrightarrow \bar{t}c$. We find that the resonance
produce cross section $\sigma(b\bar{b})$ is very small and in the
range of $10^{-1}fb-10^{-3}fb$ in most of the parameter space.
Furthermore, the backgrounds of the final state $b\bar{b}$ are
very large,  the value of $\frac{S}{\sqrt{B}}$ is smaller than 0.2
in all of the parameter space. Thus, it is very difficult to
detect the neutral top-pion $\pi^{0}_{t}$ via this process.
However, for the process $\mu^{+}\mu^{-}\longrightarrow
\pi^{0}_{t}\longrightarrow \bar{t}c$, it is not this case. As long
as the neutral top-pion mass $m_{\pi_{t}}$ is below the $t\bar{t}$
threshold, the $\pi^{0}_{t}$ can generate 36 and up to 649
observable $\bar{t}c$ events at the FMC with
$\sqrt{s}=200GeV-500GeV$. Even we take the parameter
$\varepsilon=0.08$ and $m_{\pi_{t}}=350GeV$, $\pi^{0}_{t}$ also
can generate 36 observable $\bar{t}c$ events. Thus, the process
$\mu^{+}\mu^{-}\longrightarrow \pi^{0}_{t}\longrightarrow
\bar{t}c$ might give a signal which should be easy to identify.
The properties of the neutral top-pion $\pi^{0}_{t}$ can be
studied via the process $\mu^{+}\mu^{-}\longrightarrow \bar{t}c$
at a future muon collider.

\vspace{.5cm} \noindent{\bf Acknowledgments}

 This work was supported by the National Natural Science
 Foundation of China (90203005).

\end{document}